\documentclass[10pt]{iopart}
\usepackage{graphicx}

\begin{document}

\title[Fission fragment mass yield deduced from density distribution in the pre-scission configuration]{Fission fragment mass yield deduced from density distribution in the pre-scission configuration}

\author{M Warda and A Zdeb}

\address{Department of Theoretical Physics, Maria Curie-Sk\l odowska University, Lublin, Poland}
\ead{warda@kft.umcs.lublin.pl}
\vspace{10pt}
\begin{indented}
\item[]February 2014
\end{indented}

\begin{abstract}
Static self-consistent methods usually allow to determine the most probable fission fragments mass asymmetry. 
We have applied random neck rupture mechanism to the nuclei in the configuration at the end of fission paths. Fission fragment mass distributions have been deduced from the pre-scission nuclear density distribution obtained from the self-consistent calculations.
Potential energy surfaces as well as nuclear shapes have been calculated in the fully microscopic theory, namely the constrained Hartree-Fock-Bogolubov model with the effective Gogny D1S density-dependent interaction. 
The method has been applied for analysis of fission of $^{256,258}$Fm, $^{252}$Cf and $^{180}$Hg and compared with the experimental data.
\end{abstract}

%\maketitle
\ioptwocol

\section{Introduction}
Fission is one of the dominant decay channels of  the heaviest nuclei.  Physics of this process is crucial in determining stability of heavy and super-heavy isotopes. One of the basic observables of fission, obtained directly in experiment, is fragment mass distribution. Measured yields allow to determine type of fission as well as to deduce the properties of the mother nucleus. Prediction of the charge, mass and total kinetic energy (TKE) distributions of fission fragments is still a challenging task for the theory of nuclear fission. A number of attempts have been made to describe mechanism of sharing  nucleons between fission fragments since the spontaneous fission phenomenon was discovered.

Historically, in the first theoretical description of fission the macroscopic liquid drop model was used. In this approach the competition between Coulomb repulsion and surface energy of deformed charged liquid drop of nuclear matter is analyzed. The mass distributions calculated within this method are symmetric, as a consequence of ignoring microscopic effects, that are responsible for octupole deformation of mother nucleus and deformations of the fragments~\cite{NS65}. The more sophisticated approach, scission point model~\cite{W76}, assumes that fission properties can be derived from the scission point configuration that is well defined in the evolution of fissioning nucleus. The scission point is described as a configuration of two deformed touching fragments. The potential energy is calculated using macroscopic-microscopic model. The probability of certain fragmentation is in inverse proportion to the scission point potential energy. This method allows to determine  the probability, that the certain number of nucleons will be incorporated to one of the fragments from the energy of the scission configuration. 

Within improved scission point model \cite{A04, A05, A13} one can obtain distribution of fission fragments and the mean value of total kinetic energy. In comparison to Ref.~\cite{W76} this model allows to calculate energy of interaction between fragments and deformation energy at the scission point configuration, what results on obtained mass and TKE distributions. The main disadvantage of this treatment is large number of phenomenological parameters, e.g.  touching distance of fissioning system, surface tension coefficient with parameters fitted to the magic nuclei.

The  authors of the microscopic scission point method \cite{PAN12} also deduce fission properties as fragment mass and TKE from the analysis of dinuclear system that may be created after scission. Energies of nuclei are obtained in the microscopic self-consistent calculations. Depending on mass asymmetry and deformation of the fragments energy of the system is calculated and fission probability is determined. In this approach strong assumptions on distance between fragments and deformation of the fragments at scission point have to be made.

Another approach includes  dynamics of fission in the analysis. It treats the nuclear shape evolution as a Brownian motions of nuclear system on the potential energy surface  (PES) \cite{JR1, JR2}.  The direction of the motion in the five-dimensional deformation space is determined randomly using Metropolis' method. The considered PES is calculated within macroscopic-microscopic model, where the potential energy consists of liquid drop part (with deformation-dependent coefficients) and microscopic corrections. During ``random walk" on this surface each shape might be obtained with accordance to its statistical weight. The dimensional depletion causes that agreement with experimental mass distribution is getting worse. The additional parameter -- critical neck radius was introduced, and its value results much on the obtained fragment mass yields.

There were also several attempts to describe fragment mass distribution in a fully  microscopic way, i.e. using Adiabatic Time-Dependent Hartree-Fock-Bogolubov (ATDHFB) method \cite{HG2005} or Hartree-Fock-Bogolubov (HFB) method with Skyrme Energy Functional~\cite{Schunck}.
ATDHFB model allows to include dynamical effects to the description of the fission process. The probability of mass division is proportional to the total flux of the wave function through the scission point in a particular configuration. Including dynamic effects gives  much better agreement with data in comparison with static calculations. Especially the broadness of fragment mass distribution is closer to the experimental one.

The HFB method with Skyrme energy functional was applied for description of induced fission process \cite{Schunck}. The authors discussed the impact of the triaxiality and discontinuity of the PES in the scission point region on the fragment mass distribution. Quasi-particle occupations in the nascent fragments were localized and attributed to each fragment. However, no comparison of theoretical and experimental distributions were performed.

Dynamic calculations at low excitation energy were performed using Langevin approach \cite{AC13}. Shell and pairing effects, as well as dissipation and fluctuation were included in the model. Out of fragment mass distribution it was also possible to obtain the time scale of fission process. The authors made a strong assumption, that both nascent fragments have the same deformation, what argues with several observations of shapes of fission products \cite{J}.

Published lastly GEF code \cite{KHS} predicts surprisingly good fission fragment mass yields with a very simple construction of the PES of fissioning nucleus. Namely, macroscopic potential is modified by parabolic corrections to simulate shell effects. Once fitted it reproduces wide range of experimental fission results.

The main inspiration for our investigations is the idea proposed by U. Brosa \etal \cite{B83, B88, B90}. The authors assume direct dependence of fragment mass distribution on the random neck rupture mechanism of mother nucleus in its pre-scission shape.  The method describes splitting of a nucleus as a consequence of hydrodynamic fluctuations induced by random vibrations of nuclear surface. The pre-scission deformation of the fissioning nucleus has a decisive influence on the fragment mass yield. The probability, that the rupture of the neck occurs in a certain position, decreases with the radius of a neck. In this method no assumption on properties of fission fragments is needed. This simple macroscopic model, allows to describe many fission properties as the TKE distribution and the dependence of neutron multiplicity on fragment masses. 

 We apply the HFB model, employing finite-range Gogny forces to calculate the PES and mass distribution during fission. The previous investigations performed with this model  proven, that it describes with reasonable accuracy the experimental data of the most relevant properties of unstable nuclei for different decay modes, such as spontaneous fission \cite{WERP,WE12,BGG84,DGD08} and exotic process of cluster radioactivity \cite{WR11}. Also the asymmetric fission of $^{180}$Hg isotope was successfully explained within microscopic approach \cite{WSN12}. In this paper detailed analysis of the scission point configuration determined in the microscopic calculations is performed. From the single mass distribution we  deduce possible fragment mass asymmetries.

This paper is built as follows: In the second section details of calculations are described. Results obtained for four isotopes are presented in the third section. Finaly fourth section include our conclusions and  discussion of the results.

%%%%%%%%%%%%%%%%%%%%%%
\section{Method}%
%%%%%%%%%%%%%%%%%%%%%%
The analysis of the scission point configurations requires realistic description of the nuclear matter mass distributions at large deformations. They may be provided by 
the calculations performed within the self-consistent constrained
HFB model with the effective Gogny density-dependent interaction \cite{GD80}. The popular D1S parametrization \cite{BGD91}  was used. The computer code of \cite{ERC97} was applied for numerical calculations. 
The PES was determined in the deformation space of quadrupole and octupole deformations. In the PES
the fission paths were determined as continuous lines connecting local minima of energy for fixed value of quadrupole moment. In each point of the constrained calculations nuclear matter spatial distribution can be easily derrived from the single-nucleon wave functions. 

\subsection{The pre-scission point configuration}

The scission point is defined as the configuration of nuclear system in which molecular shape (i.e. two fragments connected by  a thin neck) of a mother nucleus converts into two separate fragments. In some simple models it is represented by the configuration of two touching nascent nuclei. In fact, when a realistic leptodermous density distribution is taken into account, one can find, that during the splitting nuclear matter density in the neck  decreases gradually from the bulk value to zero while surfaces of the  fragments overlap. Thus we can not define one scission configuration but rather a scssion region in which two fragments are being separated.

\begin{figure}[th]
 \includegraphics[height=1\columnwidth,angle=270]{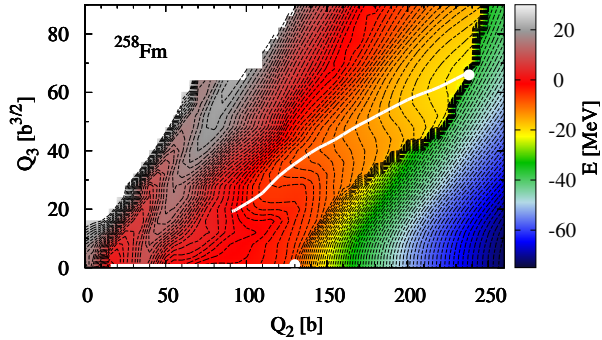}
\caption{The PES of $^{258}$Fm. The white lines correspond to the fission paths.}
\label{PES}
\end{figure}

In the self-consistent calculations of the PES it is easy to locate the ranges of the scission region. Increasing constraint on deformation parameter, e.g. quadrupole moment, leads to more elongated shape of a nucleus with a thinner neck. At some point sudden change take place. The energy minimization procedure leads not to molecular shape but a solution with two nuclei separated by a few fm distance is found.  Such dinuclear system usually has got energy much lower than the compound nucleus calculated in the previous step. Surely, scission take place between these two configurations. In this way the last point on the fission valley before rupture should be called pre-scission point and the next point with two separated fragments post-scission point. The line connecting pre-scission points on the PES should be called scission line (or more precisely pre-sisccion line), see Fig. \ref{PES}. Intermediate configurations of the scission region are hard to obtain in the self-consistent procedure as they require multiple constraints, including e.g. neck parameter \cite{WR11, WSN12}.

The evolution of shape along fission path goes smoothly with increase of deformation up to the the pre-scissin point. It was found \cite{WERP, WSN12} that many fission properties are determined by the shape  of the PES from saddle to scission, where a nucleus takes  a molecular shape. Especially the configuration in the pre-scission point is crucial in the fission process as no further evolution of a compound nucleus is allowed. 
  
In most cases determination of the scission line is trivial, as in the constrained calculations many observables essentially change their values: e.g. energy, hexadecapole moment, neck thickness. Such line can be easily  noticed in Fig.  \ref{PES} between $Q_2= 130$ b and 250 b as rapid change of energy of the system take place along scission line.

\begin{figure}[th]
 \includegraphics[width=1\columnwidth,angle=0]{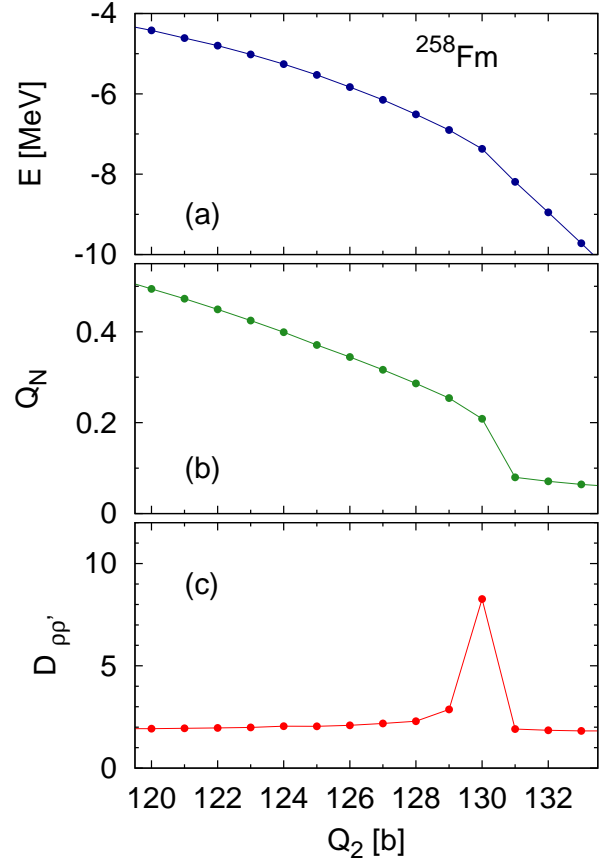}
\caption{The fragment of the compact fission path of $^{258}$Fm in the vicinity of scission point. (a) The energy, (b) the neck parameter $Q_N$ and (c) the density distance function $D_{\rho \rho'}$ between two neighbouring configurations as a function of the quadrupole moment $Q_2$.}
\label{fig1}
\end{figure}

In the case of symmetric fission path in $^{258}$Fm at $Q_3=0\, \rm b^{3/2}$ it is not so clear as molecular shape converts quite smoothly into dinuclear system without sudden drop of energy. It can be easily found in the panel (a) of Fig. \ref{fig1}, where fission path smoothly converts into hyperbolic decrease characteristic for Coulomb repulsion between separated fragments. Thus fall of energy  can not be a criterion to determine the scission point. Detailed calculations show only a tiny kink with change of energy derivative when the neck is about to disappear.  To prove, that this is real scission point, first we calculate neck parameter $Q_N$ defined as:

\begin{equation}
Q_N=\int \rho(z,r_\bot)\; \exp{\left(\frac{z^2}{a_0^2}\right)} dz dr_\bot \;
\end{equation}
where $a_0$ is a parameter describing width of a neck region with an arbitrary chosen value $a_0= 1$ fm. $Q_N$ describes number of particles in the neck region. Middle panel of Fig. \ref{fig1} shows changes of a neck parameter along the fission path. In the first part of fission path $Q_N$ decreases slowly, whereas beyond $Q_2=129$ b it drops down rapidly. It means that sudden change in density distribution take place here. Neck parameter takes small positive and almost constant values from  $Q_2=130$ b. This is characteristic behaviour of $Q_N$ for two separated nuclei when only tails of density distribution of both fragments contribute to $Q_N$.

As it was mentioned, scission point may be defined as a configuration, where rapid change of matter distribution from molecular to dinuclear system occurs. To check discontinuity in changes of density of nucleons it is useful to follow procedure of Ref.~\cite{D12}. We may define the density distance function $D_{\rho\rho^,}$  given by:

\begin{equation}
D_{\rho\rho^,}=\int \left|\rho(\mathbf{r})-\rho'(\mathbf{r})\right| d\mathbf{r},
%D_{\rho\rho^,}=\int d\mathbf{r}\; \left|\rho(\vec r)-\rho'(\vec r)\right|,
\label{D_rho}
\end{equation}
where $\rho(\bf r)$ and $\rho'(\bf r)$ are the the local spatial densities, calculated in two following steps during the nuclear shape evolution. This function takes small values during smooth evolution of nuclear shape. When two different configurations are compared it is  manifested by a sharp single peak.

In the panel (c) of Fig. \ref{fig2} we can find again that function $D_{\rho\rho^,}$ is four times  larger at $Q_2=130$ b in comparison to the neighbouring deformations. We can deduce rapid change of configuration between solutions at $Q_2=129$ b and at $Q_2=130$ b, namely rupture of the neck is observed.

All above mentioned arguments show that at $Q_2= 129$ b pre-scission point can be found on the symmetric fission path of $^{258}$Fm.

\subsection{Random neck rupture mechanism}

It was proven in Ref. \cite{WSN12, YG11} that the structure and basic properties of the nascent fragments are preliminary determined in the pre-scission configuration of fissioning nucleus. In the pre-scission point two fragments are already created in the form of two nuclei connected by  a neck. The most of nucleons are localized in the fragments, while some of them (up to 20) still create the neck. They shall be incorporated to the nascent fragments during scission. Usually it is assumed, that the neck is ruptured in its thinnest place. In this way one can calculate how nucleons are shared and the most probable mass asymmetry can be found.
Nevertheless, following the idea presented in Refs.~\cite{B88,B90}, the probability $P$ of rupture of a neck  depends on the energy $E_{cut}$ needed to create a cut in the considered position $z$ along the symmetry axis. For each position $z$ we can calculate particular fragment mass asymmetry. 

The probability $P$ of the rupture of a neck, leading to fragment mass asymmetry $A_1/A_2$ reads:
\begin{equation}
P(A_1,A_2)\sim {\rm exp}({-E_{cut}/T}),
\label{P1}
\end{equation}
where $T=\sqrt{12E^{sc}/A}$ is temperature of the pre-scission deformation. Excitation energy at scission:
\begin{equation}
E^{sc}=E_{g.s.}-E_{def}^{sc}, 
\label{Esc}
\end{equation}
which was gained during the evolution from the ground state to the pre-scission deformation. $E_{cut}$ depends on the linear density $\sigma(z)$ of a neck in a position of $z$ that corresponds to required fragment mass asymmetry. Note that two values of $z$ give the same mass asymmetry but with lighter and heavier fragment placed at opposite sides. Cut energy is defined as:

\begin{equation}
E_{cut}(z)=2\gamma \sigma(z),
\label{Ec}
\end{equation}
The expression for surface tension coefficient $\gamma=0.9517[1-1.7826(1-2Z/A)^2]$ was taken in a from given in Ref.~\cite{BRS77}. The cross section of a neck is given by:

\begin{equation}
\sigma(z)= 2 \pi \int_0^{\infty}  r_\perp \rho(z, r_\perp) dr_\perp \;,
\label{sz}
\end{equation}
where $\rho(z,r_\perp)$ is the spatial density of nuclear matter. Finally, the probability for the rupture, with corresponding division of nucleons between fragments, takes the form:
\begin{equation}
P(A_1,A_2)=\exp[-2\gamma\sigma(z)/T]
\label{P}
\end{equation}

\begin{figure}[th]
 \includegraphics[width=1\columnwidth,angle=0]{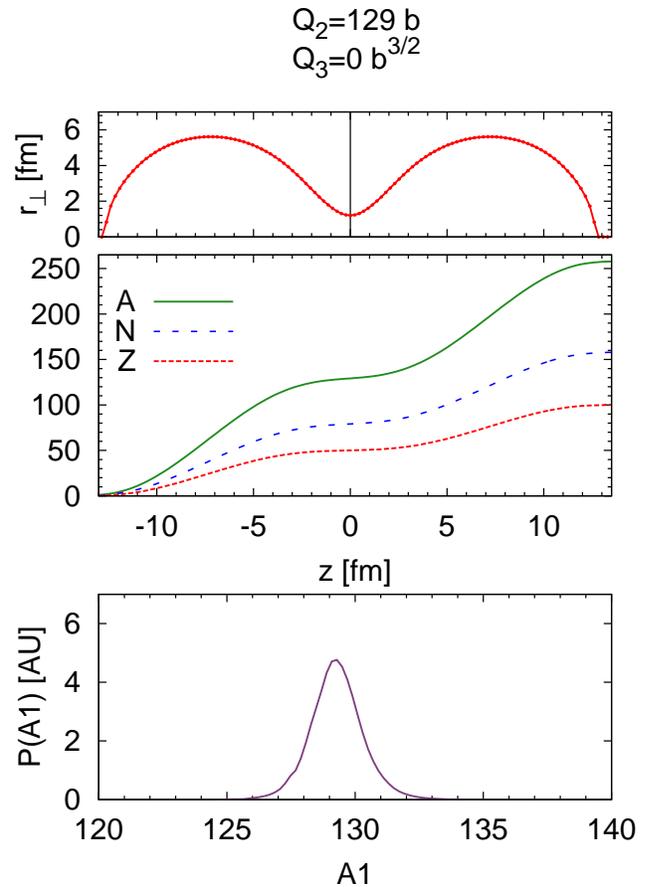}
\caption{The upper panel shows the shape of the $^{258}$Fm at a density of $\rho_0=0.08\,\rm fm^{-3}$ in the pre-scission point on the symmetric path. The number of nucleons as a function of symmetry axis $z$ (middle panel) and the probability of division of nucleons between nascent fragments, depending on the cross section of the neck in a certain position calculated using Eq.~(\ref{P}) (lower panel).}
\label{fig2}
\end{figure}

This method is visualized in Fig.~\ref{fig2}, where the case of $^{258}$Fm symmetric mode is considered. The top panel shows the shape of the nucleus at the half-bulk density $\rho=0.08$ fm$^{-3}$ in the pre-scission point \cite{WERP}. Each neck's rupture position at the $z$ axis is correlated with specific mass division between fragments shown in the middle panel of Fig. ~\ref{fig2}.  The probability of the rupture decreases strongly, while the neck becomes thicker, according to Eq. (\ref{P}). In consequence mass yield presented  in the lower panels of Fig.~\ref{fig2} is quite narrow  with the most probable position of a rupture corresponding to the equal split into two $^{129}$Sn isotopes. 

%%%%%%%%%%%%%%%%%%%%%%%%%%%%%%%%%%%%%%%%%%%%%%%%%%%%%%%%%%%%%%%%%%%%%%%%%
\section{Results}
%%%%%%%%%%%%%%%%%%%%%%%%%%%%%%%%%%%%%%%%%%%%%%%%%%%%%%%%%%%%%%%%%%%%%%%%%

\begin{figure}[th]
\includegraphics[width=0.8\columnwidth,angle=0]{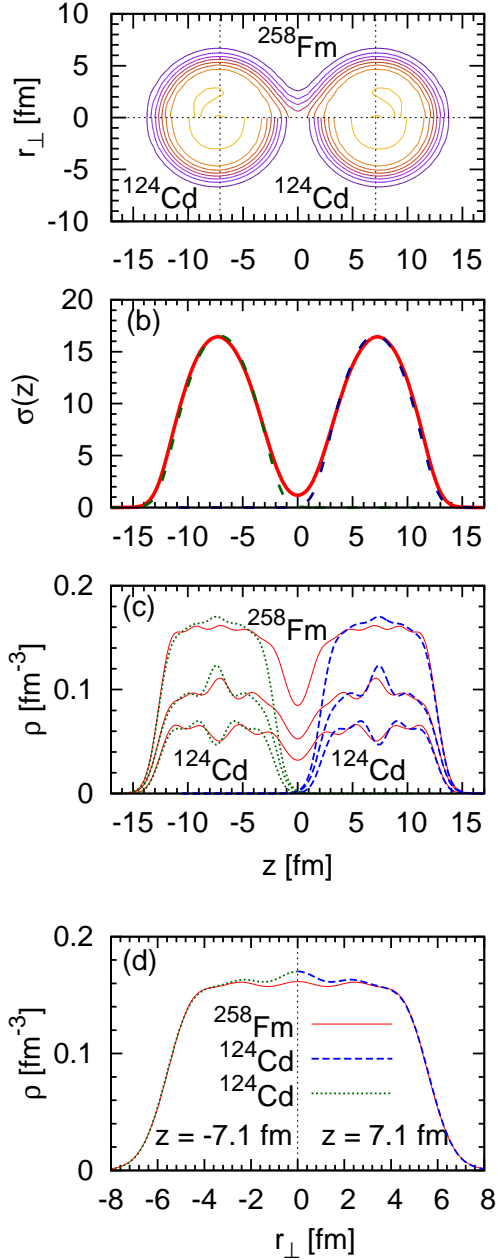}
\caption{(a) Density distribution of $^{258}$Fm in configuration close to the scission on the symmetric fission path ($Q_{20}=129$ b, $Q_{30}=0\;\mathrm{b}^{3/2}$) compared with the density distribution of $^{124}$Cd (spherical shape). (b) Linear density of $^{258}$Fm in its pre-scission shape (continuous line) and $^{124}$Cd (dotted lines). Profiles of density for cross sections taken in (c) $r_\perp=0$ fm and (d) $z=-7.1$~fm, $z=7.1$ fm.}
\label{fig3a}
\end{figure}

It was shown, that fission in $^{258}$Fm represents specific, bimodal character \cite{H86, H89}. Its mass yield is very narrow with a single peak. The TKE distribution is compounded of the low and high energy modes with equal abundance. They can be linked to the theoretically described asymmetric and symmetric fission channels on the PES, respectively \cite{WERP}, which are visible on the PES in Fig.~\ref{PES}. A nucleus in its ground state is quadrupole deformed ($Q_{20}=16$ b). The symmetric path goes along reflection symmetric shapes and terminates at deformation close to $Q_{20}=129$~b, as it was discussed in the previous Section. The density distribution corresponding to this pre-scission configuration is shown in the upper part of Fig.~\ref{fig3a}a. The shape of a nucleus can be described as a molecular system of two spherical $^{124}$Cd isotopes connected by a thin neck containing 4 protons and 6 neutrons. The fragments conserve $N/Z$ ratio of mother nucleus and reproduce the shape and mass of the outer part of the fragments before scission. In panels (c) and (d) of Fig.~\ref{fig3a} density profiles of $^{258}$Fm and two $^{124}$Cd  are compared along $z$ axis and along perpendicular axis at the thickest part of the fragments.  Out of the neck region very good agreement can be noticed.

\begin{figure}[th]
\includegraphics[angle=0,width=0.8\columnwidth]{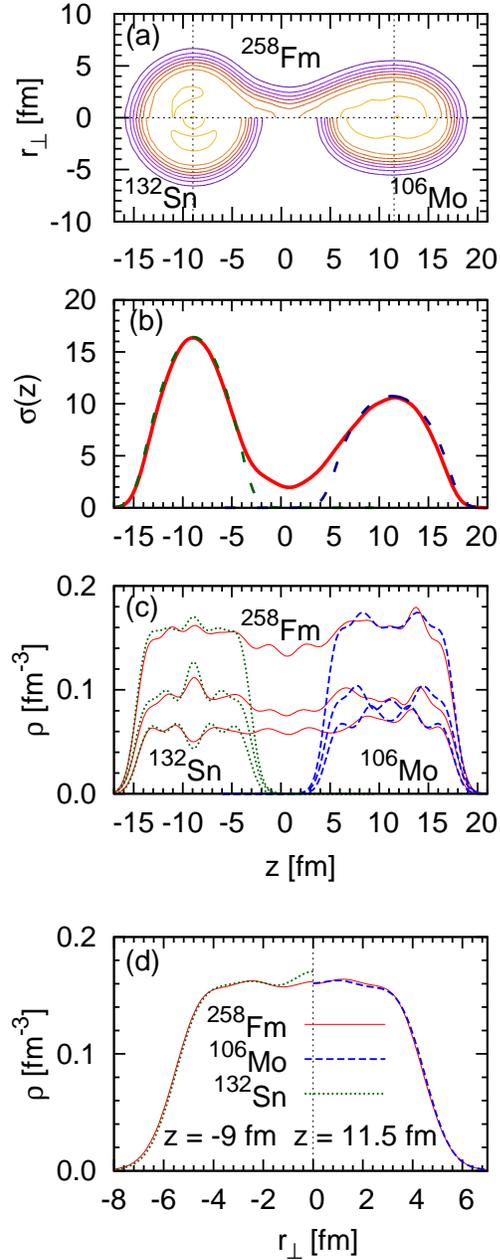}
\caption{Similar as in Fig. \ref{fig3a} but for pre-scission configuration on asymmetric path ($Q_{20}=238\;\mathrm{b}$, $Q_{30}=66.7\;\mathrm{b}^{3/2}$). Density distribution and profiles are compared with $^{132}$Sn (in its spherical ground state) and $^{106}$Mo ($Q_{20}=6\;\mathrm{b}$).}
\label{fig3b}
\end{figure}

\begin{figure}[th]
 \includegraphics[height=1\columnwidth,angle=270]{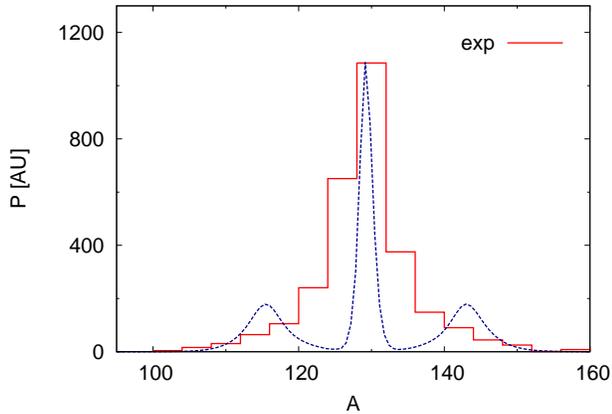}
\caption{Fragment mass distribution for the spontaneous fission of $^{258}$Fm obtained within presented method. The sum of distributions (dotted line), corresponding to the symmetric and asymmetric paths is compared with the experimental yield, taken from  Ref.~\cite{H86}.}
\label{fig5}
\end{figure}

The alternative asymmetric path arises behind the barrier from $Q_{20}\approx 90\;\mathrm{b}$ and $Q_{30}\approx 20\;\mathrm{b}^{3/2}$ and leads towards  configuration close to $Q_{20}=239$~b, $Q_{30}=67\;\mathrm{b}^{3/2}$. The analysis of density profile of asymmetric pre-scission deformation is shown in Fig.~\ref{fig3b}a. This molecular configuration consists of spherical double-magic $^{132}$Sn and prolate $^{106}$Mo. In this case, a neck contains 8 protons and 12 neutrons. Again, in panels (c) and (d) of Fig.~\ref{fig3b}, nice reproduction of the density profiles of mother nucleus by  nascent fragments can be found. 

In the symmetric mode 10 neck nucleons should be shared between fragments in the scission point. The most probable split of symmetric pre-scission shape leads to the production of two $^{129}$Sn isotopes. In the asymmetric fission channel $^{143}$Cs  and  $^{115}$Rh isotopes are most likely created. If we assume, that neck does not have to be cut in the thinnest place, these numbers would convert into  stretched fragment mass distribution. 
To this end the procedure presented in the previous section is applied twice, both  for symmetric and asymmetric pre-scission shapes. From the density distribution linear density is calculated in each case, what can be seen in Figs.~\ref{fig3a}b and~\ref{fig3b}b. Next,  fission probability as a function of mass asymmetry is calculated from Eq.~(\ref{P}). The results are combined with assumption of equal abundance of both modes and plotted in Fig.~\ref{fig5} by dotted line. In the same Figure experimental results taken from Ref. \cite{H89} are shown.   One can find, that symmetric peak is much more narrow than the experimental one.  Asymmetric mode produce two additional side  peaks at $A_H\approx 143,\;A_L\approx 115$ which are not visible in the experimental data. Nevertheless asymmetric peaks fit in the region of the tail of mass yield of $^{258}$Fm.

The main reason of inconsistency is lack of dynamic effects in the present analysis. In the final part of evolution of a nucleus, the PES is quite soft in $Q_3$ direction and wide fission valleys are developed (see Fig. \ref{PES}). We have considered only two pre-scission configuration at the end of fission paths. However, there is a big chance, that nucleus evolves to the other pre-scission configurations. Exact probability of reaching each point along scission line on the PES should be obtained in dynamical calculations. From each pre-scission shape fragment mass distribution can be deduced. Combination of these two effects would lead to broader mass distribution.

\begin{figure}[th]
\includegraphics[angle=270,width=0.8\columnwidth]{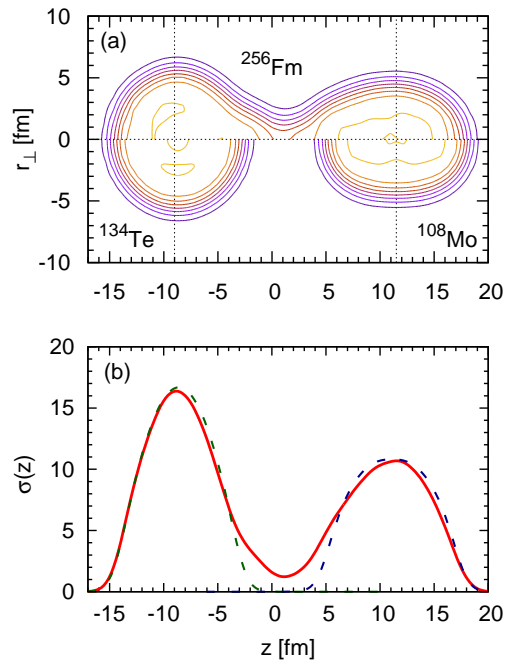}
\caption{Similar as in Fig. \ref{fig3a} but for pre-scission configuration of $^{256}$Fm ($Q_{20}=230\,\rm b$, $Q_{30}=68\,\rm b^{3/2}$). Density distribution and profiles are compared with $^{134}$Te ($Q_{20}=1\,\rm b$) and $^{108}$Mo ($Q_{20}=6\,\rm b$).}
\label{fig3c}
\end{figure}

\begin{figure}[th]
 \includegraphics[height=1\columnwidth,angle=270]{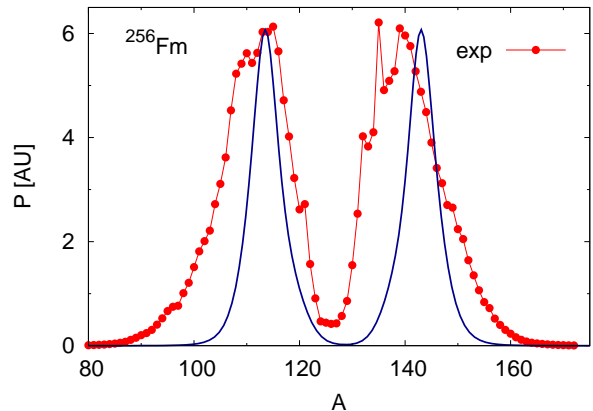}
\caption{Fragment mass distribution for the spontaneous fission of $^{256}$Fm isotope in comparison to the experimental data, taken from Ref. \cite{ENDF}.}
\label{fig6}
\end{figure}
 
The fission properties of $^{256}$Fm differ substantially from its neighbour isotope with two more neutrons. Its spontaneous fission half-life is seven orders of magnitude longer and symmetric fission channel is not active. The detailed explanation of these facts can be found in Ref.~\cite{WERP}.  In $^{256}$Fm only the asymmetric channel of fission is visible in the experiment. Analysis of the density profile of the pre-scission configuration (Fig.~\ref{fig3c}) indicates, that the nascent fragments are close to the $^{134}$Te and $^{108}$Mo with 6 protons and 8 neutrons in a neck region. From the linear density function $\sigma(z)$ presented in Fig. \ref{fig3c}b the fission fragment mass distribution are obtained within presented method. The results are shown in Fig.~\ref{fig6} together with experimental mass yields. The most probable mass of light fragment is reproduced with good accuracy. The heavy fragment peak is slightly shifted in comparison to the experimental one. The broadness of calculated distribution is reduced.

\begin{figure}[th]
\includegraphics[angle=270,width=0.8\columnwidth]{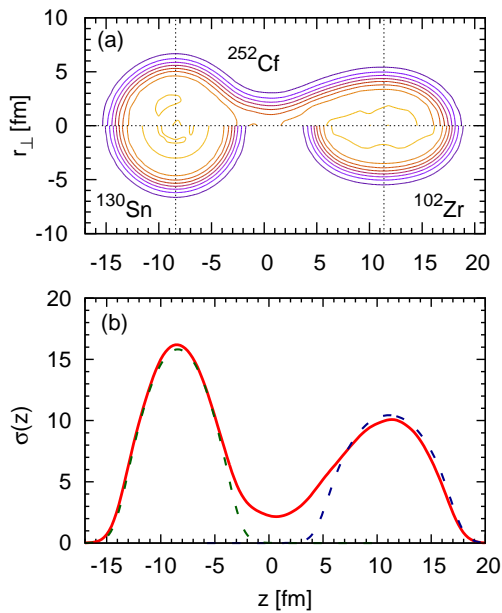}
\caption{Similar as in Fig. \ref{fig3a} but for pre-scission configuration of $^{252}$Cf ($Q_{20}=217\,\rm b$, $Q_{30}=66.5\,\rm b^{3/2}$). Density distribution and profiles are compared with $^{102}$Zr ($Q_{20}=5.5\,\rm b$) and $^{130}$Sn (spherical).}
\label{fig3d}
\end{figure}

\begin{figure}[th]
 \includegraphics[height=1\columnwidth,angle=270]{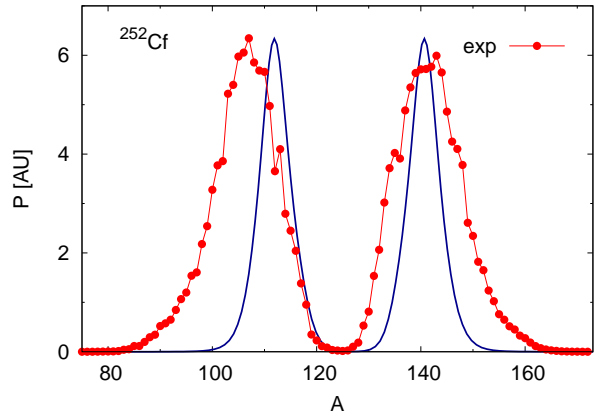}
\caption{The same as in Fig. \ref{fig6} but for $^{252}$Cf isotope. Experimental data were taken from Ref. \cite{CF}.}
\label{fig7}
\end{figure}

In Fig.~\ref{fig3d} the density profile of $^{252}$Cf in its pre-scission deformation is shown as well as its linear density function. The fragment mass distribution is presented in Fig.~\ref{fig7}. In this case, the most abundantly produced fission light fragment is not reproduced within presented method. The predicted mass distribution of the lighter fragment is shifted towards heavier masses in comparison to the experimental data. Probable explanation of these discrepancies comes from the fact, that post neutron emission measurements are presented  \cite{ENDF} and average neutron multiplicity is larger in the region of the most probable lighter fragment \cite{ZHO11}. Similar to the results of $^{256}$Fm, theoretical mass  distribution is too narrow, but the peak-to-valley ratio is in good agreement to the observed yield.

\begin{figure}[th]
\includegraphics[angle=270,width=0.8\columnwidth]{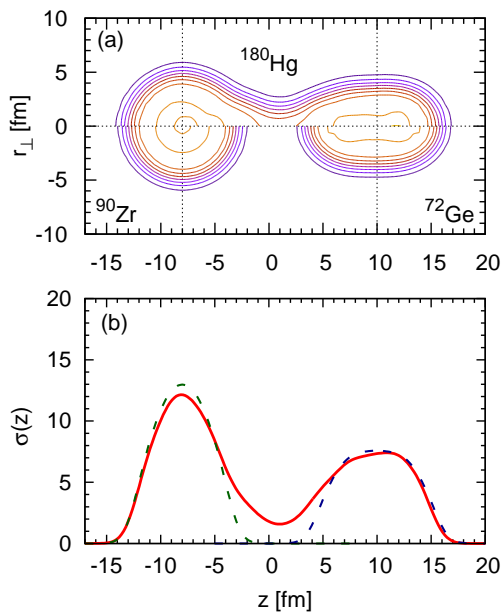}
\caption{Similar as in Fig. \ref{fig3a} but for pre-scission configuration of $^{180}$Hg ($Q_{20}=130\,\rm b$, $Q_{30}=30\,\rm b^{3/2}$). Density distribution and profiles are compared with $^{90}$Zr (spherical) and $^{72}$Ge ($Q_{20}=8\,\rm b$).}
\label{fig3e}
\end{figure}

\begin{figure}[th]
 \includegraphics[height=1\columnwidth,angle=270]{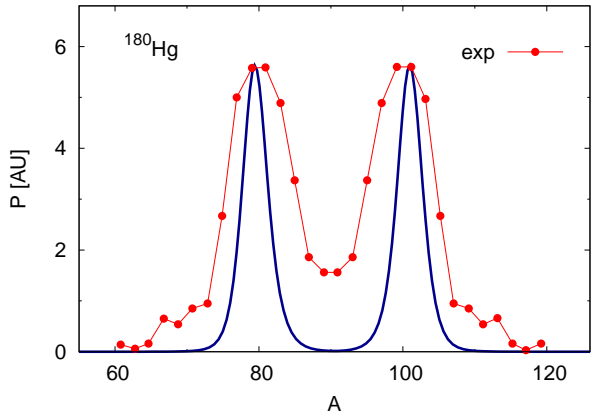}
\caption{The same as in Fig. \ref{fig6} but for $^{180}$Hg isotope. Experimental data were taken from Ref. \cite{els13}.}
\label{fig8}
\end{figure}

The recent experimental studies brought unexpected observations of $\beta$-delayed fission of $^{180}$Hg \cite{Andr10,els13}. The most abundantly produced fission fragments have got fragment mass asymmetry $A_H/A_L=100/80$ instead of expected fragmentation leading to two magic $^{90}$Zr isotopes. Detailed microscopic analysis \cite{WSN12} explained the dominant character of asymmetric fission valley. The $^{180}$Hg, as a product of electron capture in $^{180}$Tl, is created with excitation energy not larger than $10.44$~MeV. In this nucleus the pre-scission point is in the vicinity of a saddle around 12 MeV above the ground state. Scission point excitation energy should be taken equal maximal energy available for the nucleus in its ground state. Density distribution in the pre-scission point as well as function $\sigma(z)$ are plotted in Fig. \ref{fig3e}. Fragment mass distribution is compared with experimental data in Fig.~\ref{fig8}. The most probable heavy and light masses ($A_H=100$ and $A_L=80$) are well reproduced. Calculated yield is too narrow in comparison with the observed one. The peak-to-valley ratio is not reproduced as well.

%%%%%%%%%%%%%%%%%%%%%%%%%%%%%%%%%%%%%%%%%%%%%%%%%%%%%%%%%%%%%%
\section{Summary}
%%%%%%%%%%%%%%%%%%%%%%%%%%%%%%%%%%%%%%%%%%%%%%%%%%%%%%%%%%%%%%

The presented results confirm, that microscopic description of the pre-scission configuration provides a lot of information about physics of fission. Applying macroscopic method proposed by Brosa to the self-consistently  calculated  nuclear matter density distribution, we have deduced ambiguous fragment mass yield. The major characteristics of experimentally measured mass distributions, i.e.  the most probable masses of heavy and light fragment are reproduced with reasonable accuracy.

 The broadness of yields is too narrow as dynamics of the process was omitted in current investigation. The value of temperature used in our studies may also affect the results. Between pre- and post-scission configuration energy usually  significantly decreases. In consequence one may assume larger values of excitation energy and temperature, which leads to broadening peaks of mass distribution. 
In the neck rupture mechanism we have also ignored quantal properties of nucleons, which may impact on fragment mass distribution. We have shown, that even in asymmetric fission one of the fragments is close to spherical double-magic nucleus. Strong shell effects in the fragments as well as quantal nature of the neck nucleons may modify macroscopic model of random neck rupture.
 All presented distributions, deduced from our static calculations, are much narrower in comparison with experimental ones.  Further investigations require inclusion of dynamical effects. It would allow to reproduce required broadness of mass yields.

%\section{Acknowledgements}
\ack
This work is partially supported by National Science Center in Poland by grant No. 2013/11/B/ST2/04087.

\end{document}